\documentclass[12pt,aps]{article}
\usepackage{amssymb}
\usepackage{amsmath}
\newcommand{\ket}[1]{{|#1\rangle}}
\newcommand{\bra}[1]{{\langle#1|}}

\newcommand\be{\begin{equation}}
\newcommand\ee{\end{equation}}
\newcommand\bea{\begin{eqnarray}}
\newcommand\eea{\end{eqnarray}}

\begin{document}
\noindent{\large \sf The Physical Interpretation of {\sf PT}-invariant Potentials}\\ \\
\noindent{\sc \normalsize Stefan Weigert} \\ \\
\noindent{\normalsize \rm $\bra{\mbox{Hu}}\mbox{MP}\rangle$ -
\rm Department of Mathematics, University of Hull\\
UK-Hull HU6 7RX, United Kingdom \\
{\tt S.Weigert@hull.ac.uk}\\ \\
July 2004 \\ \\
\parbox[height]{13cm}{\small {\sf Abstract}: A purely imaginary potential can provide a phenomenological description of creation and absorption of quantum mechanical particles. $PT$-invariance of such a potential ensures that the non-unitary phenomena occur in a balanced manner. In spite of wells and sinks which locally violate the conservation of quantum probability, there is no net get loss or gain of particles. This, in turn, is intuitively consistent with real energy eigenvalues.} \\ \\ 

A non-unitary time evolution generated by a non-hermitean operator is deemed an undesirable if not unacceptable feature of a fundamental theory such as quantum mechanics---energy or particles numbers are not conserved. When $PT$-symmetric, non-hermitean Hamiltonians \cite{bender+98} generally tend to behave somewhat better: invariance under the combined action of a hermitean involution such as parity $\hat P$ and an anti-linear operation such as time reversal $\hat T$ may lead to a real spectrum of energy eigenvalues and to a unitary time evolution, at least in a modified sense \cite{mostafazadeh03}.

These and related observations have triggered many studies of $PT$-invariant systems \cite{czechjournal}. It turns out that to relax hermiticity to $PT$-invariance leads to natural generalizations of various concepts and physical systems. To mention only a few: supersymmetry, exactly and quasi-exactly solvable models, perturbation theory, random-matrix ensembles, field theory,  periodic potentials, and scattering theory have seen the birth of their $PT$-invariant twins \cite{PTtopics}. In a sense, many of these developments are straightforward since they are based on replacing a real by an imaginary ``coupling constant,'' and on imposing an additional symmetry condition. These modifications do not spoil most of the calculations which have been carried out successfully for hermitean operators; often only minor changes are required to redo the calculations.     

Interesting as they are, these developments have one feature in common which, by an outsider, might be considered as a fundamental flaw: the {\em physical interpretation} of $PT$-symmetric systems remains obscure. It is the purpose of the present paper to point to a simple but fundamental feature of $PT$-invariant potentials which, possibly, may help to answer this criticism. It is argued that purely imaginary, antisymmetric potentials describe a situation in which sources and sinks for quantum mechanical probability are distributed in a {\em balanced} manner. The resulting phenomenological interpretation of $PT$-invariant potentials might---or might not---provide some comfort to the physically inclined mind which longs to give meaning to otherwise only formal manipulations.                        

Students of quantum mechanics may have come across non-real potentials in textbooks \cite{gottfried89}. More generally, non-hermitean expressions have been used to phenomenologically describe absorptive optical media, inelastic scattering from nuclei, or other loss mechanisms on the atomic or molecular level \cite{nonherm uses}; more recently, particle physics rediscovers their potential usefulness \cite{PTparticles}.  

Let us illustrate by means of an elementary example that non-real potentials are indeed capable to model both absorption and  emission. Consider a quantum particle with mass $m$ on the real line, described by Schr\"odinger's equation,
\be \label{schroe}
i \hbar \frac{\partial \ket{\psi}}{\partial t} 
 = \hat H \ket{\psi} \, , 
\ee
where the Hamiltonian $\hat H$ is given by 
\be \label{ham}
\hat H = \frac{\hat p^2}{2m} + v (\hat x ) + i w (\hat x) \, , 
 \quad v(x), w(x) \in \mathbb{R} \, .
\ee
To see that the imaginary part of the potential acts as a source or a sink for quantum particles, let us introduce two purely imaginary potentials differing only by a minus-sign, 
\be \label{complexdelta}  
W_\pm (x) \equiv i w_\pm(x) 
          = \pm i w_0 \delta (x) \, , \quad w_0 > 0 \, .
\ee
It is easy to verify that the functions 
\be \label{solution}
\psi_\pm (x) = \left\{ 
\begin{array}{rr}
e^{ \mp i k x} & \quad x \leq 0 \, , \\ 
e^{ \pm i k x} & \quad x \geq 0 \, ,
\end{array} \right. \quad k= mw_0/\hbar^2 \, ,
\ee
are solutions (with real energy $E_0=mw_0^2/2\hbar^2$) of the time-independent Schr\"odinger equation associated with (\ref{schroe}), for the potentials $W_\pm (x)$, respectively. The solutions are continuous at the origin and they satisfy the matching conditions imposed by the $\delta$-function at $x=0$. In the presence of $W_+ (x)$, the function $\psi_+ (x,t) = \psi_+ (x) \exp [-iE_0t/\hbar]$ represents a wave which travels to the left for negative values of $x$, and to the right for positive values of $x$. In other words, $\psi_+ (x,t)$ is an {\em outgoing} wave with momentum $k$. This is only possible if quantum particles are {\em created} continuously at the origin. Similarly, the solution $\psi_- (x,t)$, associated with the potential $W_-(x) = - W_+ (x)$, is readily understood as a {\em sink} for particles with energy $E_0$, streaming in at a constant rate from $\pm\infty$, only to be annihilated at the origin. Formally, the solutions $\psi_\pm (x,t)$ are closely related to the non-relativistic Green's functions for the  Schr\"odinger equation for a free particle in one spatial dimension \cite{baym73}. Details about scattering from a single $\delta$-function with complex-valued strength can be found in \cite{molinas+96}, for example.      
     
It is instructive to look at the continuity equation for the probability density $\rho (x,t) = \psi^*(x,t) \psi(x,t)$  in the position representation, 
\be \label{cont}
\frac{\partial \rho(x,t)}{\partial t} + \nabla j (x,t) 
  = \frac{2}{\hbar} w(x) \rho(x,t) \, ;
\ee
here $j(x,t) = (\hbar/2mi)(\psi^*(x,t) \nabla \psi(x,t) - \psi(x,t) \nabla \psi^*(x,t))$ is the probability flux, and the operator $\nabla$ stands for the derivative $\partial/\partial x$. It is not difficult to confirm the observations made above: for non-zero $\rho(0,t)$, the right-hand-side of (\ref{cont}) acts as a sink or a well for probability density, depending on the sign of the function $W(x)$.

Let us now turn to a non-hermitean potential which consists of two $\delta$-functions with imaginary coefficients ``of opposite sign,'' separated by a distance $\Lambda = 2\lambda$ from each other,  
\be \label{deltaPT}
W_{PT}^\lambda (x) 
           = w_\lambda i(\delta (x-\lambda) 
                                    - \delta (x+\lambda))  
           \equiv w_\lambda \delta_{PT}^\lambda (x) \, ,   
\ee
where $w_\lambda$ is a real number. The potential changes sign both under reflection at the origin, $x \to - x$, and under complex conjugation, hence, it is $PT$-invariant: $(W_{PT}^\lambda)^* (-x) = W_{PT}^\lambda (x)$. Its impact, when added to a real, symmetric potential $v_s(x)$ has been studied in \cite{PTdelta}, for example. It does not come as a surprise that one can combine the functions $\psi_\pm (x)$ to satisfy the matching conditions imposed by $\delta_{PT}^\lambda(x)$ at $x=\pm \lambda$. This leads to the following stationary solution,
\be \label{psiPT}
\psi_\lambda (x) = \left\{
\begin{array}{rrr}
e^{ + i 2 k \lambda}e^{+ i  k x} & \quad x \leq -\lambda \, , & \\ 
              e^{ - i  k x} & \quad | x | \leq \lambda \, , & \quad k = m w_\lambda/ \hbar^2 \, ,\\
e^{- i 2 k \lambda} e^{+i k x} & \quad x \geq \lambda \, ,& 
\end{array} \right.
\ee
which is an eigenstate of the operator $\hat P \hat T$:
\be \label{PTinvofpsi}
\hat P \hat T \psi_\lambda (x) = \psi_\lambda^* (-x) = \psi_\lambda (x) \, .
\ee
This is consistent with the energy eigenvalue  $E_\lambda = m w_\lambda^2 /2\hbar^2$ being real. Physically, the function $\psi_\lambda(x,t) = \psi_\lambda(x) \exp[-iE_\lambda t/\hbar]$ 
corresponds to a wave with momentum $k>0$ incident from the left, being transmitted entirely to the right.  Across the region of interaction, the wave picks up a phase shift $\delta_\lambda = -4k\lambda$. Therefore, the quantity $| \delta_\lambda/k |$ equals twice the length $\Lambda$ of the interaction region; this is clearly linked to the fact that for $|x| \leq \lambda$, $\psi_\lambda (x)$ is a wave with momentum $-k<0$, i.e. it travels in the {\em opposite} direction.  

It is interesting to briefly reflect upon the perfect transmittivity of the potential $W_{PT}^\lambda (x)$, for waves with momentum $+k$. Actually, {\em none} of the incoming particles makes it ever to the right: each incoming particle is absorbed at $x=-\lambda$; complete transparency of $W_{PT}^\lambda (x)$ is only possible since, at $x=+\lambda$, particles with energy $E_\lambda$ are being created at the appropriate rate, half of which are subsequently emitted to the right.

Qualitatively, a more general $PT$-invariant po\-ten\-tial $W_{PT} (x) = i w_a(x)$, with $w_a(-x)$ $= -w_a(x) \in \mathbb{R}$, is expected to have properties similar to those of $W_{PT}^\lambda(x)$. This follows from writing $W_{PT} (x)$ as a continuous superposition of $PT$-symmetric $\delta$-type potentials (cf. Eq. (\ref{deltaPT})): 
\be \label{PTgen}
W_{PT} (x) 
 = i \int_{-\infty}^{+\infty} d \lambda \, w_a(\lambda) 
     \delta (x-\lambda)
 = \int_{0}^\infty d \lambda \, w(\lambda) \delta_{PT}^\lambda  (x) \, .
\ee
The function $w_a(x)$ being odd ensures that particle creation and annihilation is globally balanced: in spite of  violating the conservation of probability density at each point $x$ where $w_a(x)\neq 0$ holds, the total number of particles remains constant leading to an {\em effectively} unitary time evolution compatible with real energy eigenvalues. This reasoning is also consistent with the continuity equation in the presence of a potential $W_{PT} (x)$. Spatial integration over any region symmetric with respect to the origin will make the  right-hand-side of Eq. (\ref{cont}) vanish if the probability density $\rho(x,t)$ is an even function. This, however, is guaranteed for a $PT$-invariant state such as (\ref{psiPT}) since (\ref{PTinvofpsi}) implies $\rho(x,t) = \psi(-x) \psi(x) \equiv \rho(-x,t)$. 
     
The presence of a real symmetric confining potential, $v_s(-x) = v_s(x)$, appears not to affect the inner workings of a $PT$-symmetric potential as outlined above. Therefore, the effectively unitary time evolution generated by $W_{PT}(x)$ through globally balanced emission and absorption of probability density, and the unitary flow of probability density mediated by $v_s(x)$, are expected to coexist  peacefully. From this perspective, it becomes intuitively plausible that bound states with real energy eigenvalues may emerge for an overall $PT$-invariant potential $V_{PT} (x) =v_s(x)+i w_a (x)$.

\end{document}